\documentclass[journal]{IEEEtran} 
\usepackage{cite}
\usepackage{graphicx}
\usepackage{amsmath}
\usepackage{grffile}

\graphicspath{{eps/}}

\ifCLASSOPTIONcompsoc
  \usepackage[caption=false,font=normalsize,labelfont=sf,textfont=sf]{subfig}
\else
  \usepackage[caption=false,font=footnotesize]{subfig}
\fi

\begin{document}
\title{Development of next generation LLRF control system for J-PARC rapid cycling synchrotron}
\author{Fumihiko~Tamura, Yasuyuki~Sugiyama, Masahito~Yoshii, Masatsugu~Ryoshi 
\thanks{Manuscript received June 15, 2018}
\thanks{Fumihiko~Tamura, Yasuyuki~Sugiyama and Masahito~Yoshii are with the J-PARC Center, 
JAEA \& KEK, 2-4 Shirakata, Tokai, Ibaraki 319-1195, Japan.
(e-mail: fumihiko.tamura@j-parc.jp)
}%
\thanks{Masatsugu~Ryoshi is with Mitsubishi Electric TOKKI systems Corporation, 4-11 Techno Park, Sanda, Hyogo 669-1339, Japan}
}

\maketitle
\thispagestyle{empty}
\begin{abstract}
The low level rf (LLRF) control system for the rapid cycling synchrotron
(RCS) of the Japan Proton Accelerator Research Complex (J-PARC) started
its operation in 2007. The key functions of the LLRF control system are
the dual harmonic auto voltage control, which generates superposed
voltages of the fundamental accelerating harmonic and the second
harmonic in a single wideband magnetic alloy (MA) cavity, and the
multiharmonic rf feedforward to compensate the beam loading in the MA
cavity caused by high intensity beams. These functions are necessary to
accelerate high intensity proton beams. The system has been working well
without major problems for more than ten years. However, the old FPGAs
(Xilinx Virtex-II pro etc.) are discontinued and not supported by the
current development environment. It will be difficult to maintain the
system in near future. Thus, we are planning to replace the existing
VME-based LLRF control system with a new MicroTCA.4 based system. The
system controls twelve cavities independently and calculates vector sum
of the cavity voltages in real time for phase feedback. Signal and data
transfer between the modules is a key to realize the functions. In the
existing system, the transfer is implemented not only the backplane but
also serial link via cables between the VME modules.  It is much more
simplified in the new system thanks to the high speed communication
capability of the MicroTCA.4 backplane. In this article, we describe the
configuration of the system under development, the implemented
functions, and preliminary test results. 

\end{abstract}

\begin{IEEEkeywords}
Proton Synchrotron,
Low level rf (LLRF),
field-programmable gate array (FPGA),
MicroTCA.4
\end{IEEEkeywords}


\section{Introduction}
\IEEEPARstart{T}{he} rapid cycling synchrotron (RCS) in the Japan
Proton Accelerator Research Complex (J-PARC) \cite{TDR} acts as a high intensity
proton driver, which delivers high intensity proton beams 
to the Material and Life Science Experimental Facility (MLF)
for generation of neutrons and muons, as well as the injector
for the main ring synchrotron (MR). The design output beam power
of the RCS is 1~MW.

The beam commissioning of the RCS started in October 2007 and the output
beam power has been steadily increasing with
progress of the beam tuning and hardware upgrades.
During the high intensity beam study performed in January 2015,
$8.3\times 10^{13}$ protons, which corresponds to the beam power
of 1~MW at the repetition rate of 25~Hz, were successfully
accelerated with a low beam loss below 0.2\%\cite{hotchi:prab17}. The demonstration
was performed in the single shot mode, where a beam pulse is injected
from the linac to the RCS on demand.
As of May 2018, the output beam power for the MLF is 500~kW and the RCS
delivers $6.5\times10^{13}$~ppp to the MR.

\begin{figure}[t]
 \centering
 \includegraphics[width=\linewidth]{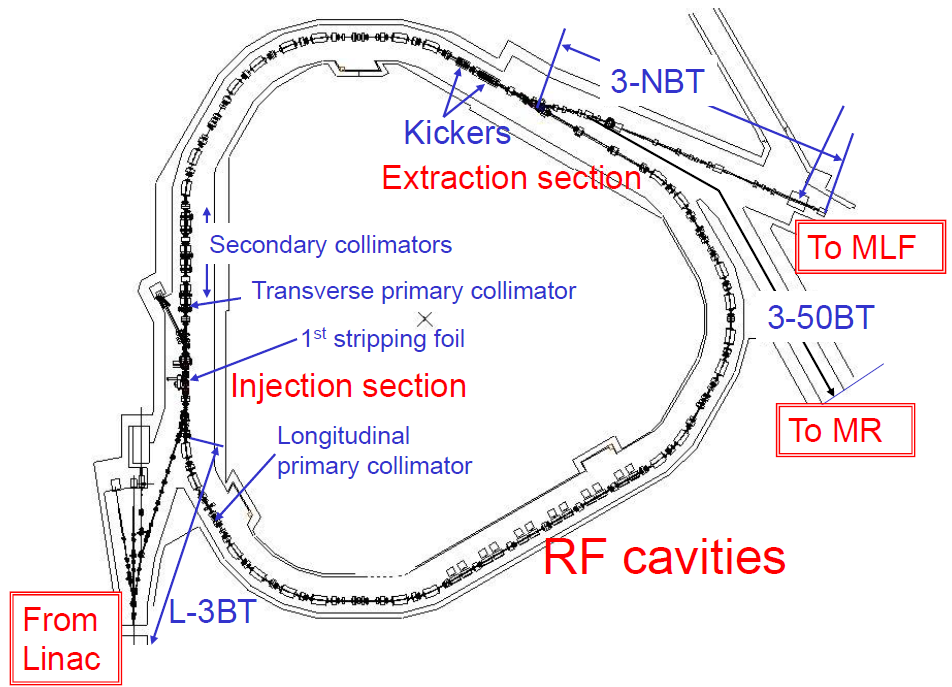}
 \caption{Schematic view of the J-PARC RCS.}
 \label{fig:layout}
\end{figure}

\begin{table}[tb]
\begin{center}
  \caption{Parameters of the J-PARC RCS and its rf system}\label{tab:parameters}
 \begin{tabular}[c]{ c c}
   \hline
   parameter &  \\ \hline
   circumference & 348.333~m \\
  energy &  0.400--3 GeV \\
   beam intensity &  (achieved) $8.3\times10^{13}$ ppp\\ 
   harmonic number & 2 \\
   accelerating frequency & 1.227--1.671~MHz \\
   maximum rf voltage &  440~kV \\
   repetition rate & 25~Hz \\
   No. of cavities &  12\\
   Q-value of rf cavity& 2 \\
   \hline
  \end{tabular}
\end{center}
\end{table}

\begin{figure*}[t] 
\centering
\includegraphics[width=0.75\textwidth]{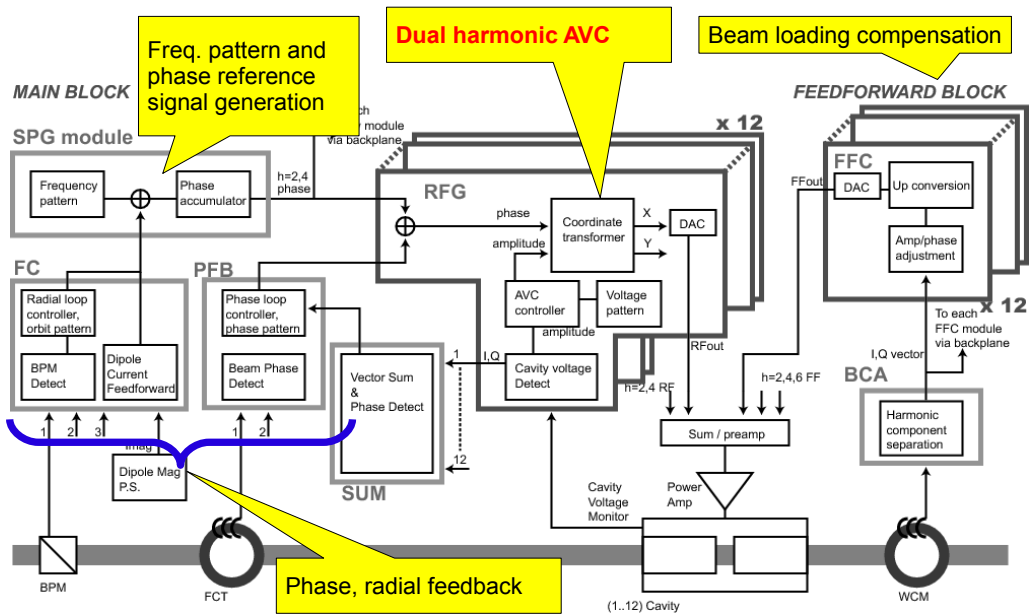}
 \caption{Block diagram of the existing LLRF control system.}
\label{fig:existing_LLRF}
\end{figure*}
\begin{figure}[t] 
\centering
\includegraphics[width=\linewidth]{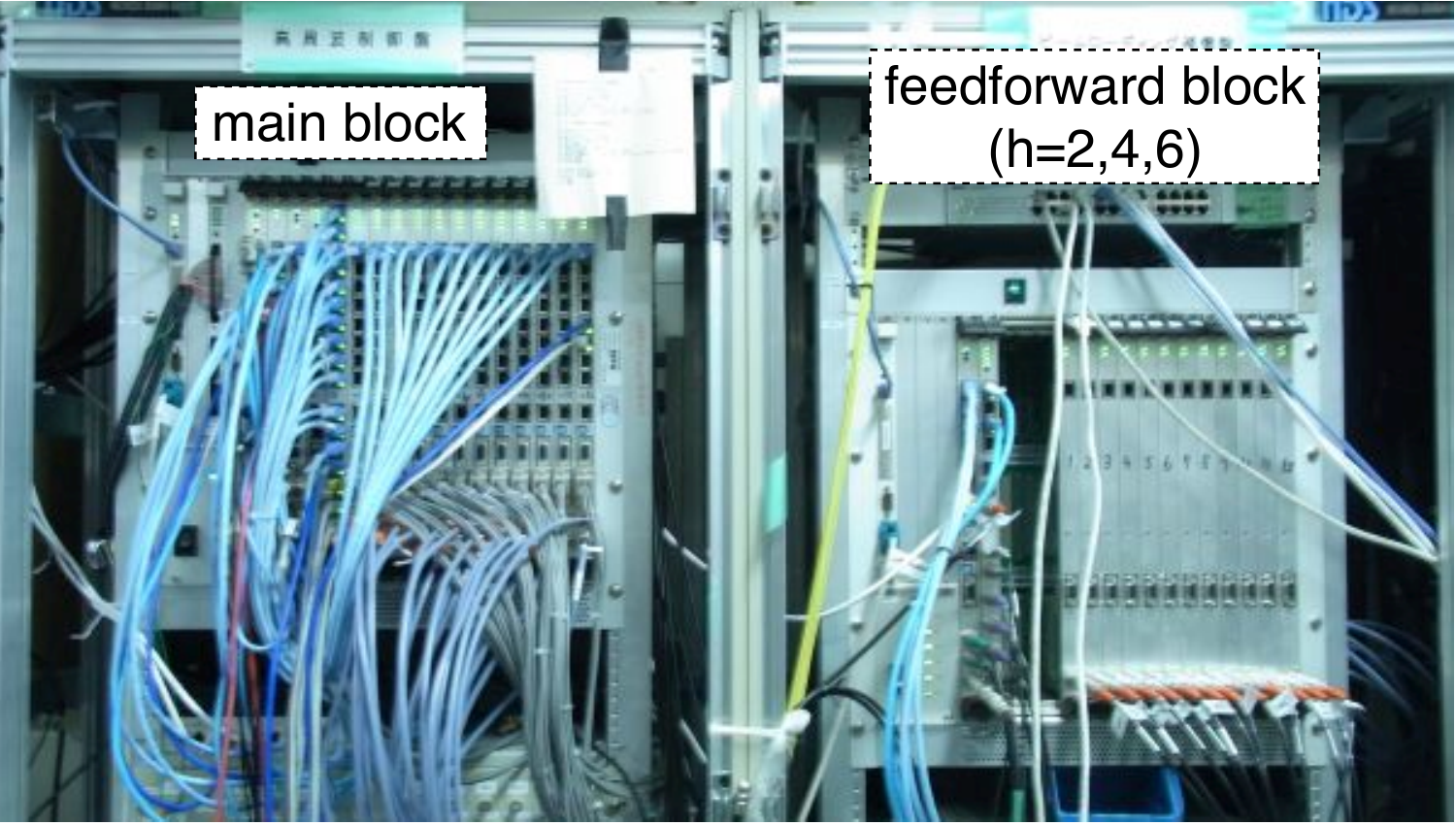}
 \caption{Photo of the existing LLRF control system.}
\label{fig:existing_LLRF_photo}
\end{figure}

A schematic view of the RCS is shown in Fig.~\ref{fig:layout}, and
the parameters of the RCS and its rf system are listed in Table~\ref{tab:parameters}.
A 400~MeV $H^{-}$ beam is injected and converted to a proton beam by
a charge exchange foil. To avoid longitudinal beam losses,
the injected beam has a chopped structure synchronized to the RCS
rf voltage. The RCS accelerates the protons up to 3~GeV in 20~ms
with the repetition rate of 25~Hz. As shown in Fig.~\ref{fig:layout},
the RCS has a three-fold symmetry. The three straight sections are
dedicated for the injection devices and the collimators, the extraction
devices, and the rf systems. 

Twelve magnetic ally (MA) cavities are installed in the RCS to
generate the high accelerating voltage of 440~kV maximum
for acceleration of high intensity proton beams.
The cavities are driven by tetrode tube amplifiers.
The MA cavity has a wideband frequency response ($Q=2$),
which covers not only the wide accelerating ($h=2$) frequency sweep 
to follow the velocity change
of the proton beam during acceleration without a tuning bias loop,
but also the frequency range of the second harmonic ($h=4$).
The wideband frequency response enables the dual harmonic operation,
where each cavity is driven by the superposition of the fundamental
and second harmonic rf voltages for bunch shaping.
The bunch shaping with the dual harmonic operation is indispensable
for alleviating the space charge effects of the high intensity proton
beams.

The beam loading in the cavity \cite{Pedersen75} is a key issue for accelerating
high intensity proton beams. In case of the wideband MA cavity,
the wake voltage consists of not only the accelerating harmonic,
but also the higher harmonics. A multiharmonic beam loading 
compensation is necessary.

These functions and other functions are implemented in the low level rf
(LLRF) control system. The LLRF control system is a key for stable
acceleration of high intensity protons.  The existing LLRF control
system started its operation in 2007 from the beginning of the beam
commissioning and operation of the J-PARC RCS.  After a decade of
operation, a next generation LLRF system for the RCS is under
development. In this article, we describe the configuration of the new
system.

\section{Existing LLRF control system}
\subsection{Configuration and functions}
The functional block diagram and the photograph are shown 
in Fig.~\ref{fig:existing_LLRF} and Fig.~\ref{fig:existing_LLRF_photo}、
respectively.
Specialized 9U height VME modules were developed to realize the functions.
The P1 connector is connected to the normal VME bus and the P2 and P3 connectors
are specialized ones dedicated for the signal transfer between the modules.
All functions are implemented as logic circuits on FPGA, Xilinx Virtex-II pro
and Spartan-2. The system clock frequency is 36~MHz.

To realize the frequency sweep, a frequency pattern memory is
implemented in a module. The revolution frequency signal is fed to a phase
accumulator to generate the revolutional phase signal from $-\pi$ to
$\pi$.  The phase signals of the higher harmonics are generated by
multiplying the revolutional phase signal by the harmonic number $h$,
therefore the synchronization between the revolutional and higher
harmonics is guaranteed. The multiharmonic phase signals are distributed
to all modules via the backplane. By using the phase signals,
synchronization of the modules for all cavities is also guaranteed.
Sinusoidal signals for the I/Q demodulation and modulation of the
rf and beam signals are generated from the phase signals.

Since the frequency range of the cavity and beam signals is relatively low, 
MHz range, no additional analog parts for down and up conversion are 
necessary; the beam or cavity signal is directly digitized by ADCs and
cavity driving signal is generated by DACs.

The main role of the system is the regulation of the
cavity voltages. Each cavity has the dual harmonic auto
voltage control (AVC) loop\cite{tamura:prst-ab08-1}.
In the dual harmonic AVC, the I/Q signals of the
accelerating ($h=2$) and the second harmonic ($h=4$) voltages
are converted to the amplitudes. The amplitudes are compared to
the voltage patterns and via PI controllers, the AVC outputs
the amplitude control signal. Finally, from the amplitudes
and the phase signals of the harmonics ($h=2,4$), the dual
harmonic rf signal is generated. The longitudinal painting
injection \cite{tamura:prst-ab09} is achieved by using the
dual harmonic AVC.

The other important function for the high intensity acceleration
is the multiharmonic beam loading compensation.
The rf feedforward method is employed in the existing system\cite{tamura:prst-ab11}.
The feedforward system picks up the complex amplitude of the
beam signal for the selected harmonic. The gain and phase
is set so that the feedforward signal cancels the wake voltage
in the cavity. We developed the commissioning methodology
of the multiharmonic feedforward system\cite{tamura:prst-ab11}.
Originally the feedforward system for the even harmonics ($h=2,4,6$)
was installed and commissioned, and the additional feedforward
system for the odd harmonics ($h=1,3,5$) was installed for
the high intensity single bunch operation\cite{tamura:prst-ab15}.

The feedback loops to stabilize the beam are also implemented.

The radial loop modulates the frequency using the beam position monitor
(BPM) signal so that the beam orbit is centered in the bending magnets.
The radial feedback is not used, because the reproducibilities of the
bending field and the rf frequency good enough.

\begin{figure}[t] 
\centering
\includegraphics[width=\linewidth]{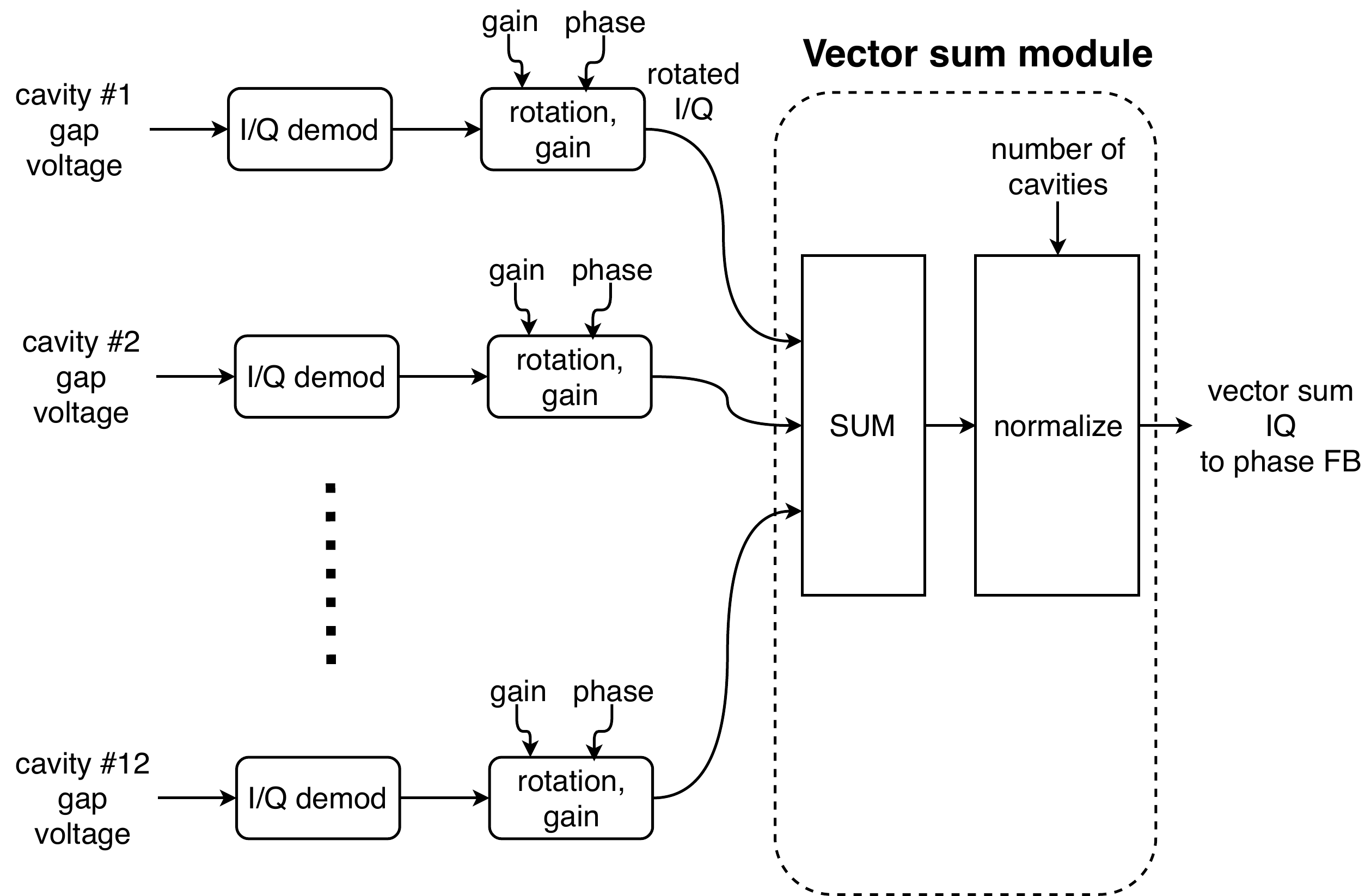} 
 \caption{Schematic diagram of the vector sum.}
\label{fig:vector_sum}
\end{figure}

The phase feedback modulates the phases of the cavity voltages
to damp the longitudinal dipole oscillation. It compares
the beam phase and the phase of the vector sum of the cavity
voltages. In Fig.~\ref{fig:vector_sum}, the block diagram
of the vector sum function is illustrated. The detected
I/Q cavity voltage of the harmonic is rotated and sent to
the vector sum module. The rotation angle is set corresponding
to the cavity position in the RCS ring. Optionally a gain
can be applied to the I/Q signal. 
The summation signal is normalized by using the number of cavities
and sent to the phase feedback module.

The miscellaneous functions not shown in the figure, 
such as generation of 
the trigger pulse for the extraction kicker magnets
and generation of the linac chopper pulse, are also
implemented.

\subsection{Demand of the next generation system}
The existing LLRF control system started its operation in 2007,
and has been working well without major problems for more than ten years.
The dual harmonic AVC, multiharmonic feedforward and the other
LLRF functions serve the high intensity beam operation.
However, the old FPGAs (Xilinx Virtex-II pro and others) used in 
the modules are already discontinued and not supported by the current
development environment. Although we have several spare modules,
it will be difficult to maintain the existing system in near future.
Therefore, we decided to develop a next generation LLRF control system.

Since we developed the existing modules by analogy of analog LLRF modules,
the module design is different for each of the functions.
Maintenance of the spare modules is a practical issue.
More generic configuration, the generic FPGA board and additional I/O board for example,
is preferable for the new system.

\begin{figure}[t] 
\centering
\includegraphics[width=\linewidth]{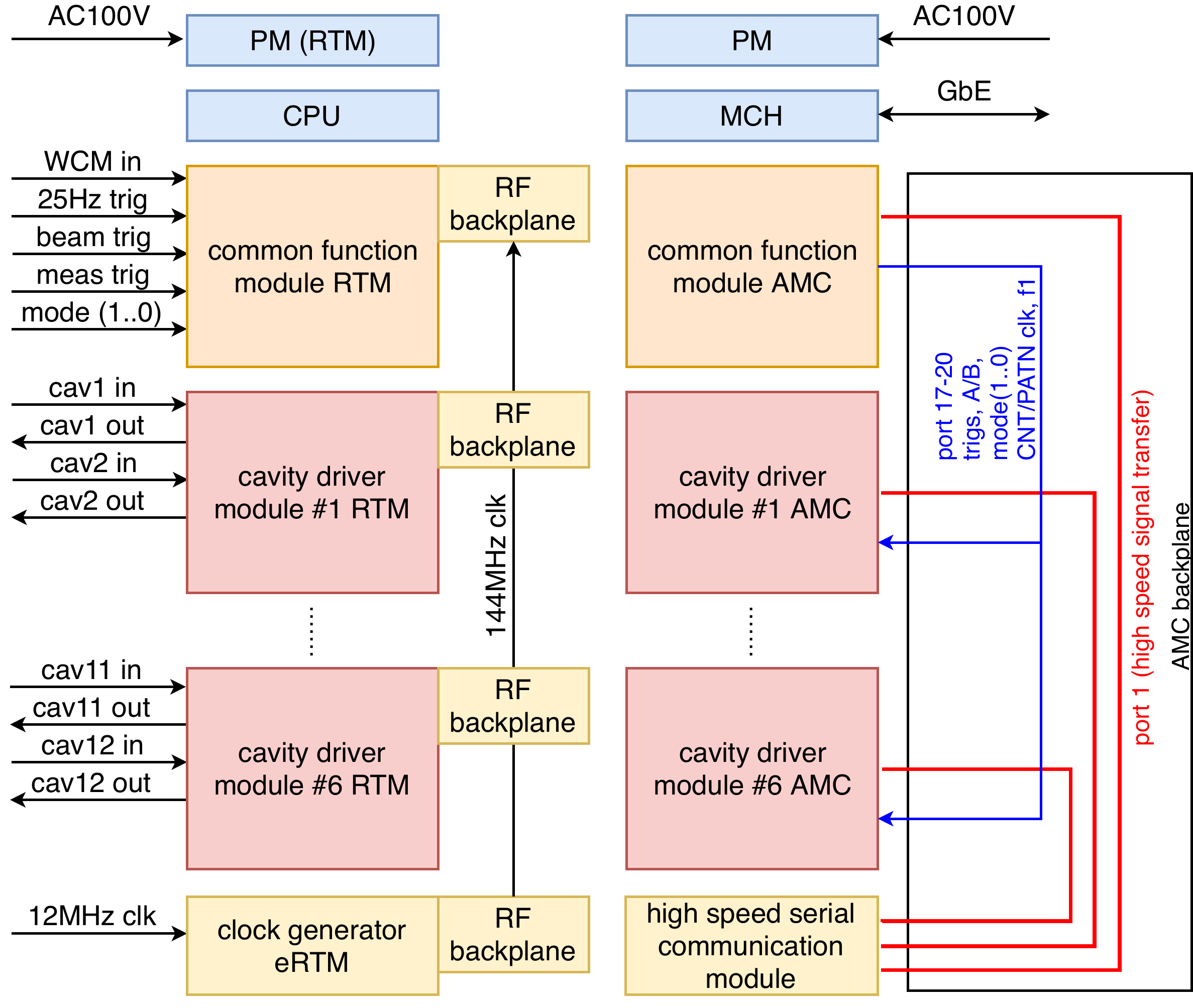} 
 \caption{Configuration of the next generation LLRF control system.}
\label{fig:new_LLRF}
\end{figure}

\begin{figure}[t] 
\centering
\includegraphics[width=\linewidth]{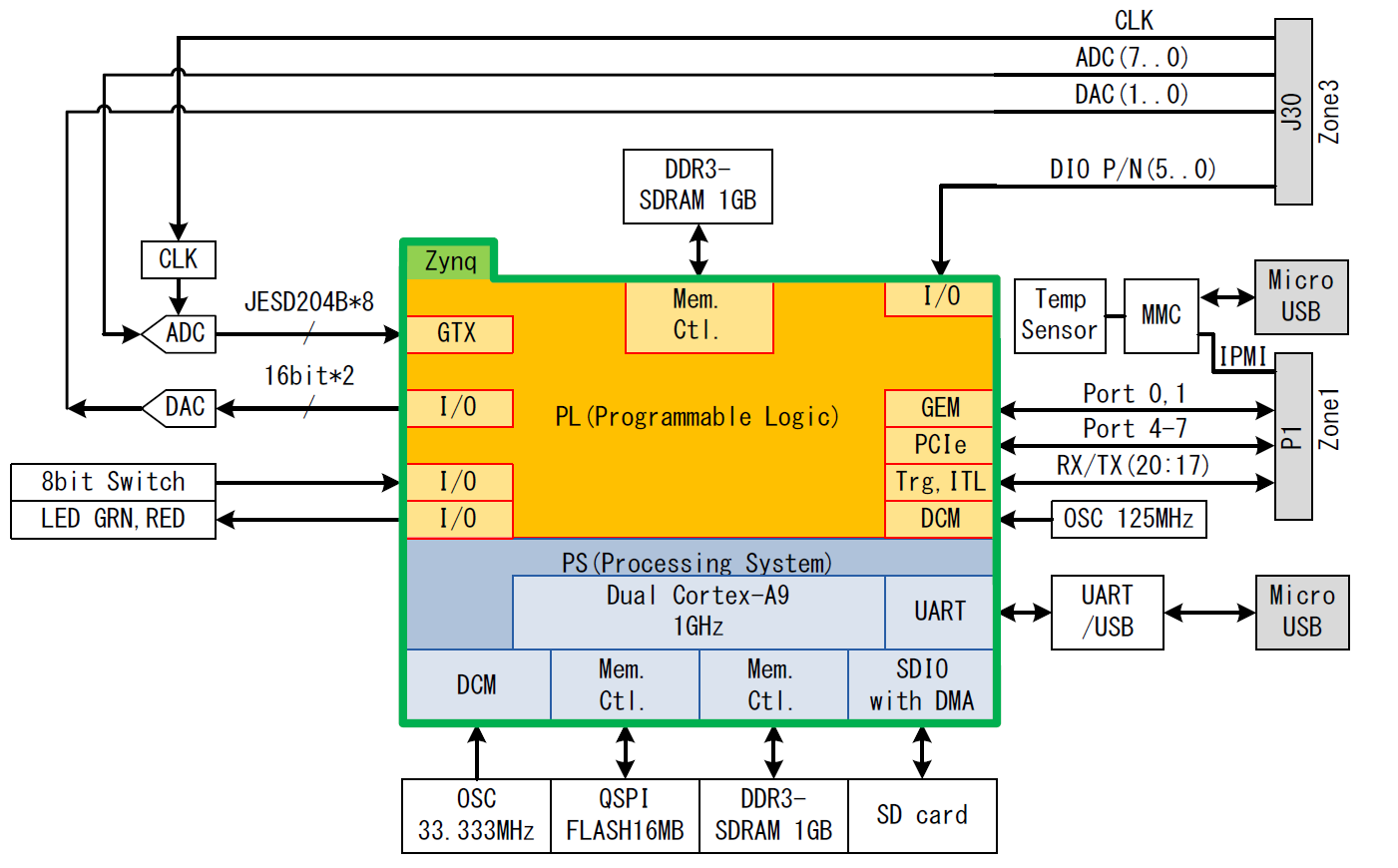} 
 \caption{Functional block diagram of the AMC module for the next generation
LLRF control system.}
\label{fig:AMC}
\end{figure}

\begin{figure}[t] 
\centering
\includegraphics[width=\linewidth]{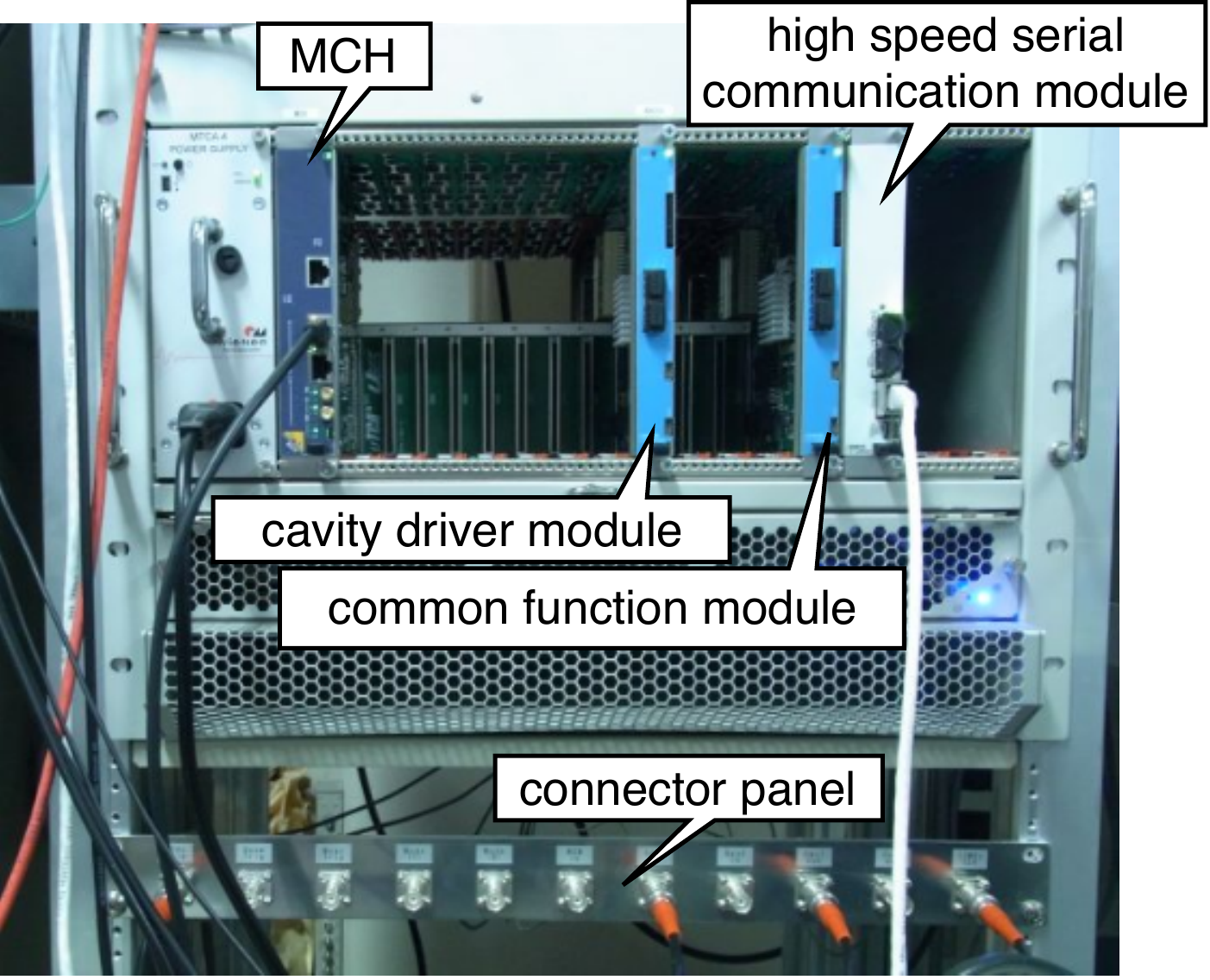} 
 \caption{Photograph of the next generation LLRF control system.}
\label{fig:new_llrf_photo}
\end{figure}

\begin{figure}[t] 
\centering
\includegraphics[width=\linewidth]{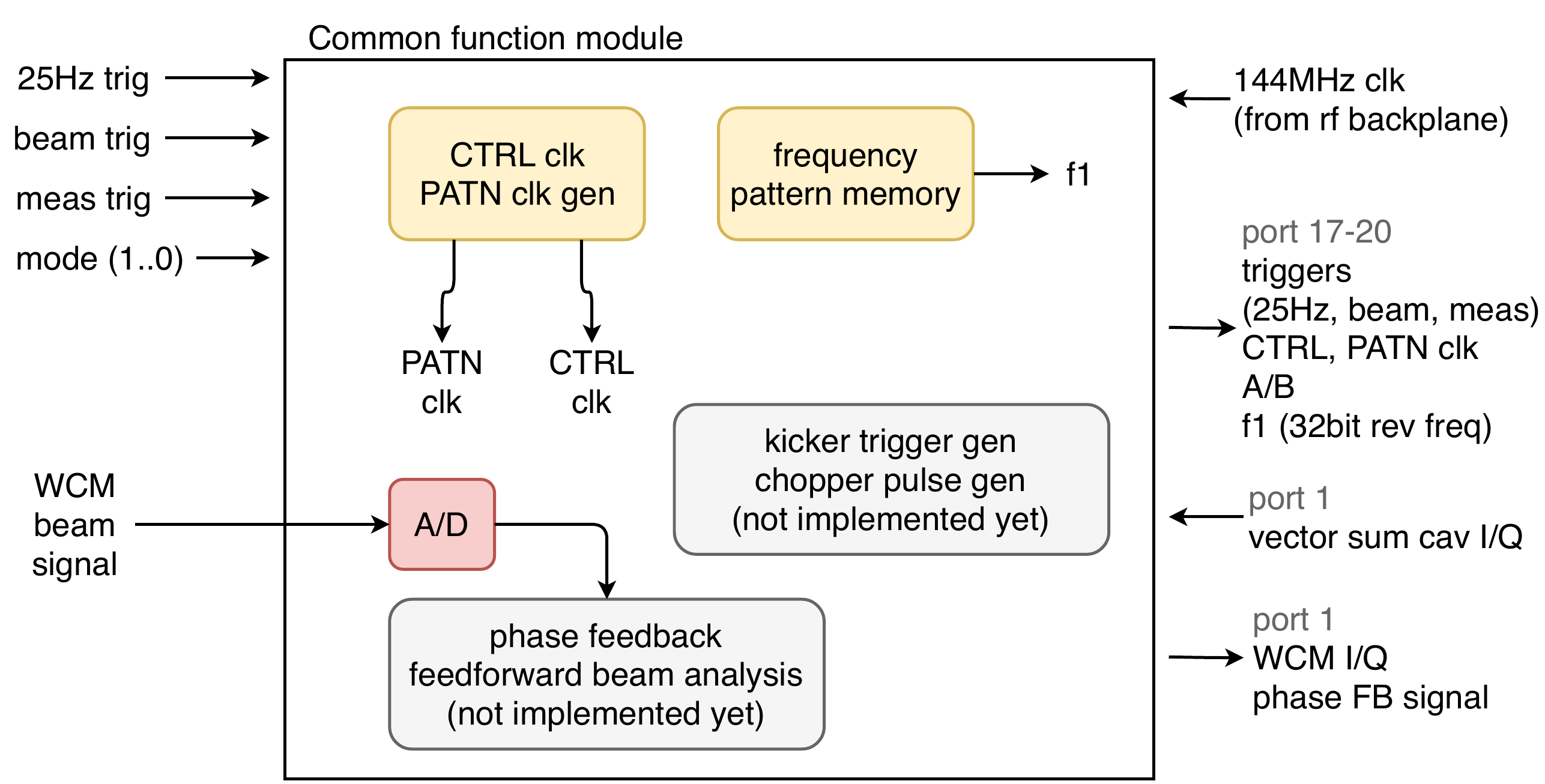} 
 \caption{Block diagram of the common function module.}
\label{fig:common_function_block}
\end{figure}

\section{Next generation LLRF control system}
\subsection{System overview}
We employ the MicroTCA.4 platform for the next generation LLRF control system.
Separation of the I/Os in rear transition modules (RTM) and the FPGA logic
in AMC modules gives us design flexibility.

The configuration of the system is shown in Fig.~\ref{fig:new_LLRF}.
The clock generator eRTM generates the 144~MHz system clock from the J-PARC
master clock of 12~MHz by using a phase lock loop.
The DESY-type rf backplane is utilized for system clock distribution
to the modules.

The general purpose AMC module
developed by Mitsubishi Electric TOKKI Systems Corporation is employed.
The block diagram of the AMC board is shown in Fig.~\ref{fig:AMC}.
It has a modern SoC FPGA, Xilinx Zynq XC7Z045, where an EPICS IOC
with Linux is embedded.  Setting and monitoring of the parameters
are done via EPICS channel access. The EPICS waveform records of I/Q signals
are useful for commissioning of the system.
The 1~GB SDRAM is used as pattern memories.
It has eight high speed ADCs and two DACs, i.e. it has capability
to control two cavities. Also, it has 6-bit digital I/O.
The RTM are developed for the specific I/O and functions.

We classify the LLRF functions into the categories, ``common function''
and ``cavity driving function'', which are implemented the common function
modules and the cavity driver modules, respectively.
A high speed serial communication module is located in the slot of MCH2.

A photograph of the next generation LLRF control system is shown in Fig.~\ref{fig:new_llrf_photo}.
In JFY 2017, one common function module, one cavity driver module, the clock generator eRTM, and
the high speed serial communication module were constructed. 

\subsection{Common function module}
A functional block diagram of the common function module is illustrated
in Fig.~\ref{fig:common_function_block}.
The common function module manages the revolutional frequency pattern, 
the phase feedback to damp
the longitudinal oscillations, and other functions.
The common function module receives the triggers and the information
of the RCS beam destination, ``mode (1..0)'', as shown in Fig.~\ref{fig:new_LLRF}.
The common function module generates the control clock used in the feedback blocks
and the pattern clock for pattern sampling. Frequencies of the clocks can be
set independently. They are set to 1~MHz for this application. 
These clocks and informations are distributed to the cavity driver modules
via the AMC backplane.
The 32-bit revolutional frequency signal from the pattern memory is serialized and distributed,
while the existing system distributes the phase signals of the accelerating ($h=2$)
and the second harmonic ($h=4$).
The cavity driver module has its own phase
accumulator and multiplier to generate the multiharmonic phase signals.
This configuration is necessary for the multiharmonic vector rf voltage
control described below.

At present, the phase feedback, the beam signal analysis for rf feedforward,
the kicker trigger generation, and the chopper pulse generation, are 
not implemented yet. The radial feedback is not to be implemented based
on our experience with the existing system.

\begin{figure}[t] 
\centering
\includegraphics[width=\linewidth]{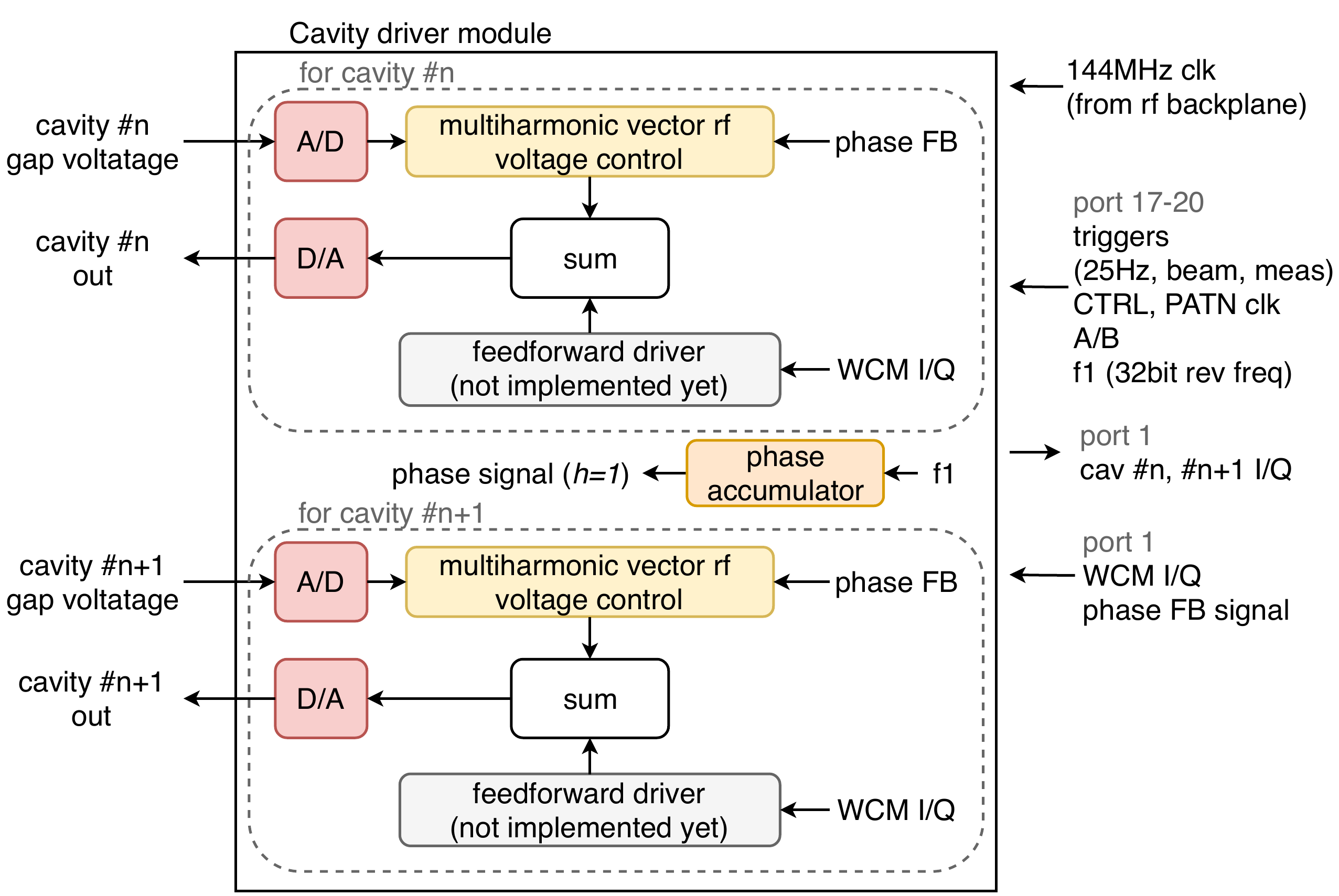} 
 \caption{Block diagram of the cavity driver module.}
\label{fig:cavity_driver_block}
\end{figure}

\begin{figure}[t] 
\centering
\includegraphics[width=\linewidth]{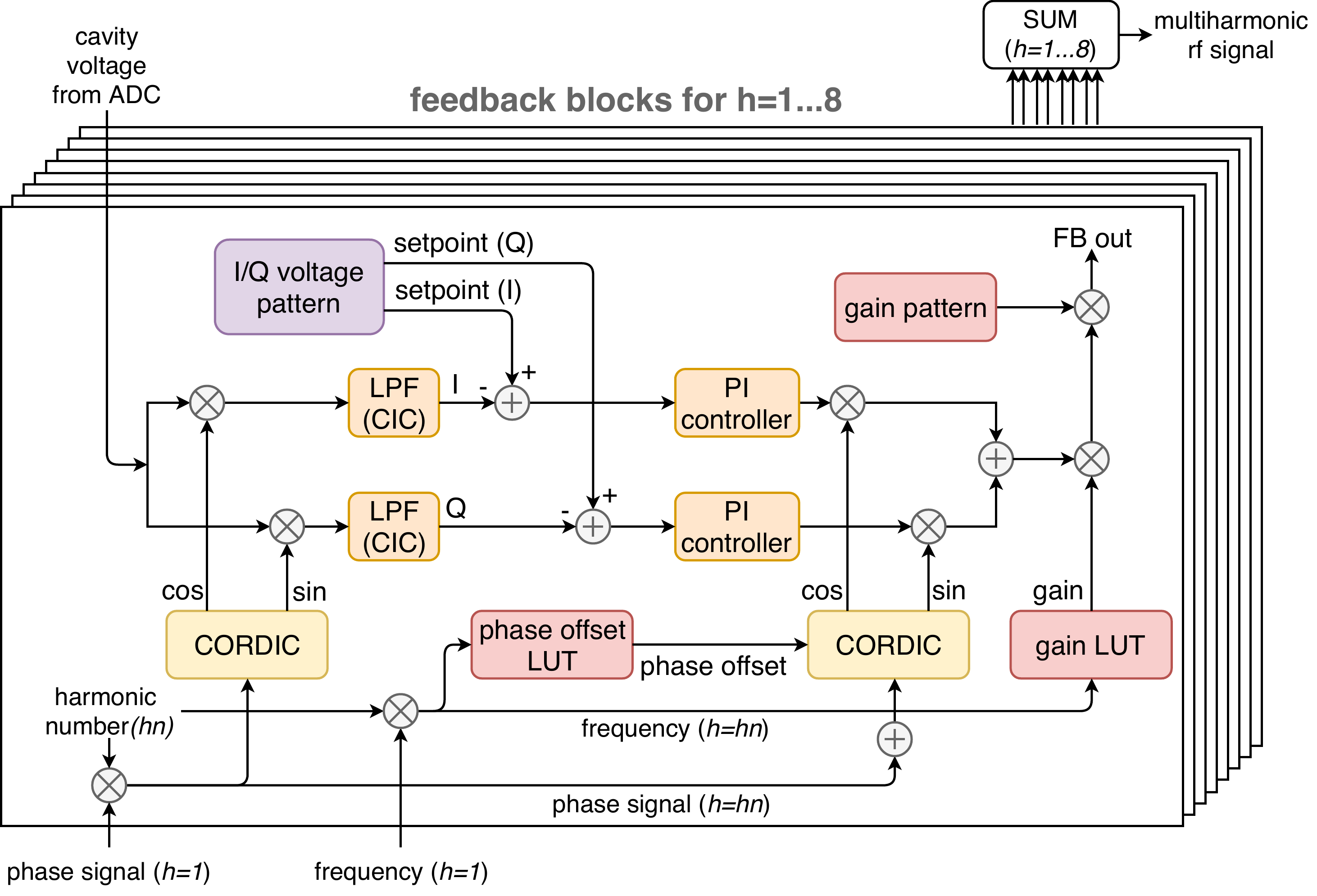} 
 \caption{Block diagram of the multiharmonic vector rf voltage control.}
\label{fig:vector_AVC}
\end{figure}

\subsection{Cavity driver module}
A functional block diagram is shown in Fig.~\ref{fig:cavity_driver_block}.
As described above, it handles two cavity voltages independently by using
two ADCs and two DACs. Six cavity driver modules are necessary to
control twelve cavity voltages, while one module is constructed in JFY
2017. 

The revolutional frequency signal from the backplane is led to the
phase accumulator to generate the phase signal, which is multiplied
by the harmonic numbers in the function blocks to generate the multiharmonic
phase signal.
The functions of the cavity driver are the multiharmonic vector rf voltage control and
the feedforward driver. Available logic cells of the Zynq FPGA
are much more than that of old FPGAs used in the existing system;
now these functions for two cavities can be implemented in a single
FPGA. 

The feedforward driver receives the I/Q amplitudes of the beam signal
for the selected harmonics and generates the feedforward compensation
signal similarly to the existing rf feedforward system.
The feedforward driver is not implemented yet. 
The number of harmonics is to be extended from six of the existing system
to eight.

The multiharmonic vector rf voltage control is the key function of the 
next generation LLRF control system. In the existing system,
the amplitudes of two harmonics ($h=2,4$) are controlled.
The new system can control the I/Q complex amplitudes of eight harmonics
($h=1...8$). By controlling the complex amplitudes, the beam loading
is compensated and the phase control of the higher harmonics is
possible.

The block consists of eight feedback blocks as shown in Fig.~\ref{fig:vector_AVC}.
The I/Q complex amplitude of the cavity voltage signal is obtained
by I/Q demodulation. A narrow band CIC (cascaded integrator and comb)
filter is used as a low pass filter. The complex amplitude is compared
to the I/Q voltage pattern. Through the PI controller and the I/Q modulator,
feedback output is obtained.

The revolutional frequency signal and the phase signal are multiplied by
the harmonic number ($hn$) in the feedback block to obtain the frequency
and the phase signal of the selected harmonic number ($hn$), respectively.
The sine and cosine signals for the I/Q demodulator and modulator are
generated by the CORDIC using the phase signal of the harmonic.
The frequency signal is used for addressing of the phase offset LUT
and the gain LUT. The phase offset LUT gives a phase offset
between the I/Q demodulator and modulator to adjust the phase
of the 1-turn transfer function. The gain LUT compensates
the amplitude response of the cavity. These LUTs are necessary to
cover the wide frequency range. 

Finally, eight rf signals from the feedback blocks are summed
up to obtain the multiharmonic rf signal.

\begin{figure}[t] 
\centering
\includegraphics[width=\linewidth]{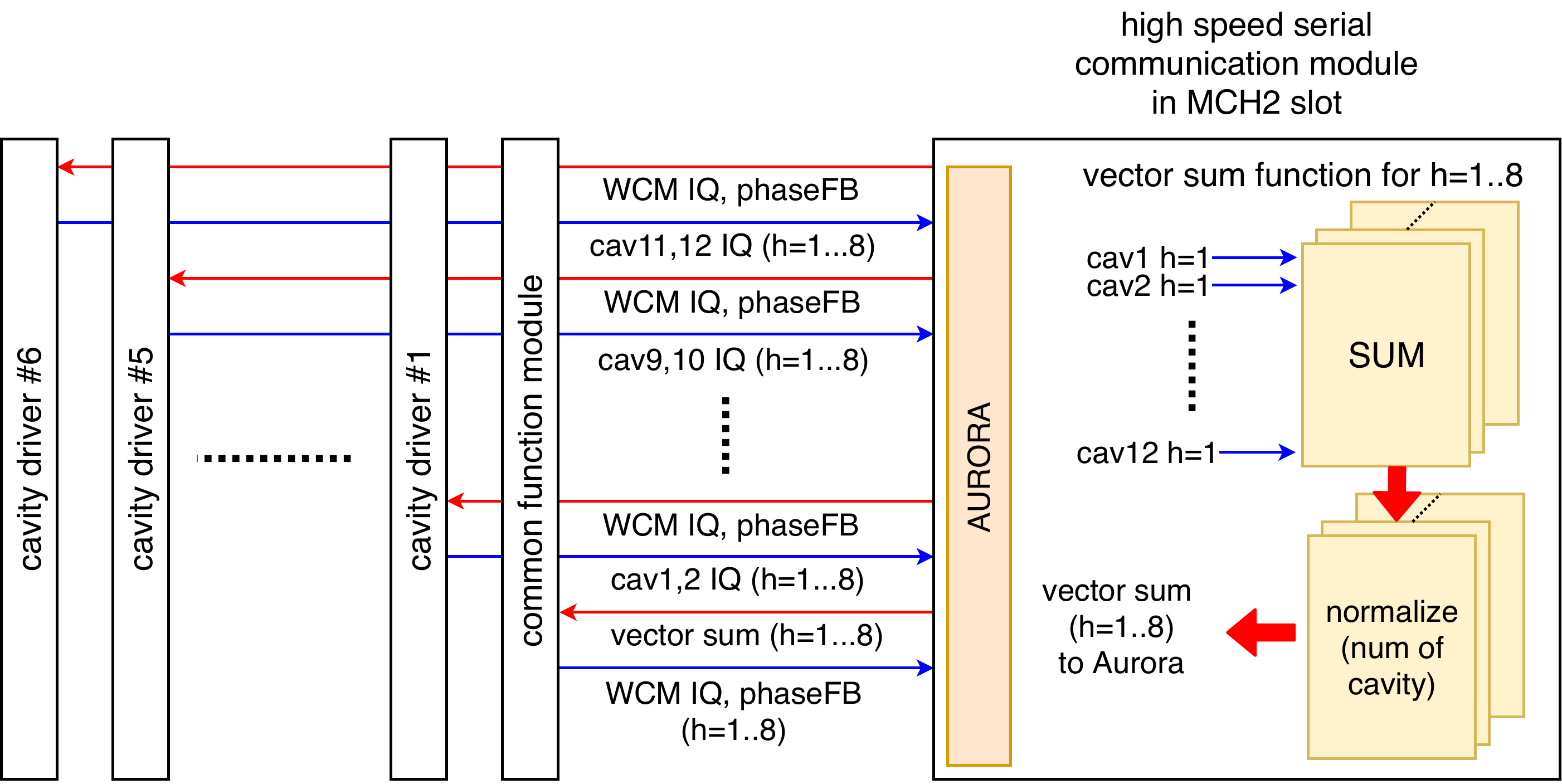} 
 \caption{Block diagram of the high speed serial communication module and the signal flow
around the modules.}
\label{fig:high_speed_serial_block}
\end{figure}

\begin{figure}[t] 
\centering
\includegraphics[width=\linewidth]{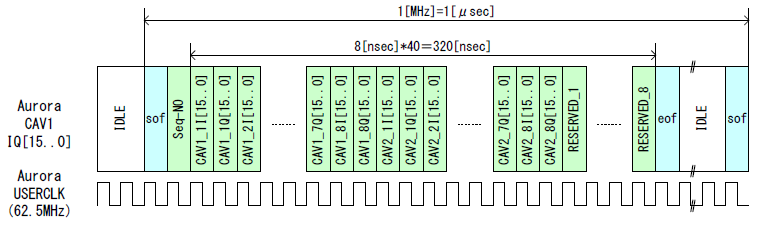} 
 \caption{Data format for the cavity I/Q signals.}
\label{fig:IQ_data_format}
\end{figure}

\subsection{High speed serial communication module}

Signal transfer between the modules is a key to realize the LLRF functions.
Actually, the signal transfer of the existing system is not very
sophisticated; parallel bus connection in the backplane is used for distribution
of the multiharmonic phase signal and the cavity I/Q voltages are sent to
the vector sum module by serial link via cables across 
the front panels of the modules, as shown
in the photograph in Fig.~\ref{fig:existing_LLRF}.

The signal transfers required by the LLRF functions are as follows.
\begin{itemize}
 \item I/Q amplitudes of the cavity voltages for the all harmonics
       from the cavity drivers to the vector sum function
 \item I/Q amplitudes of the WCM beam signal from the common function
       module to the cavity driver modules
 \item phase feedback  signal from the common function
       module to the cavity driver modules
\end{itemize}
One can see that all of the transfers are star topologies.
A star topology can be implemented by using the port 1 connections 
of the AMC backplane and installing a dedicated module in the slot
for MCH2, while the configuration sacrifices the redundancy.

The block diagram of the high speed serial communication module
and the signal flow around the module are illustrated in Fig.~\ref{fig:high_speed_serial_block}.
The cavity driver module sends the I/Q amplitudes of the two cavities for eight harmonics ($h=1..8$),
which are rotated according to the position in the tunnel, to the communication module.
To realize a number of serial connections, Xilinx Virtex-5 is employed. 
The Xilinx Aurora protocol is employed for the signal transfer.
The data format for the cavity I/Q signals is shown in Fig.~\ref{fig:IQ_data_format}. 
The data rate is set to 2.5~Gbps and a single data frame contains 40 data blocks.
Therefore, the time width of the frame is 320~ns.
In case of the I/Q signals for two cavities, 32 data blocks are actually used.
The data frame is sent every control clock cycle, 1~$\mu$s.

In the communication module, the vector sum function similar to Fig.~\ref{fig:vector_sum}.
The I/Q amplitudes from the cavity drivers are summed up and normalized by the number
of cavities. The vector sum of the all harmonics is sent to the common function
module and it is used for the phase feedback loop.

The I/Q amplitudes of the WCM beam signal for eight harmonics and the phase feedback signal
are sent from the common function module to the communication module.
The communication module distribute the signals to all cavity drivers.

Thanks to the capability of the AMC backplane for the high speed serial communication, the signal
transfer between the LLRF modules is much more sophisticated and simplified than 
the existing system.

\begin{figure}[t] 
\centering
\includegraphics[width=\linewidth]{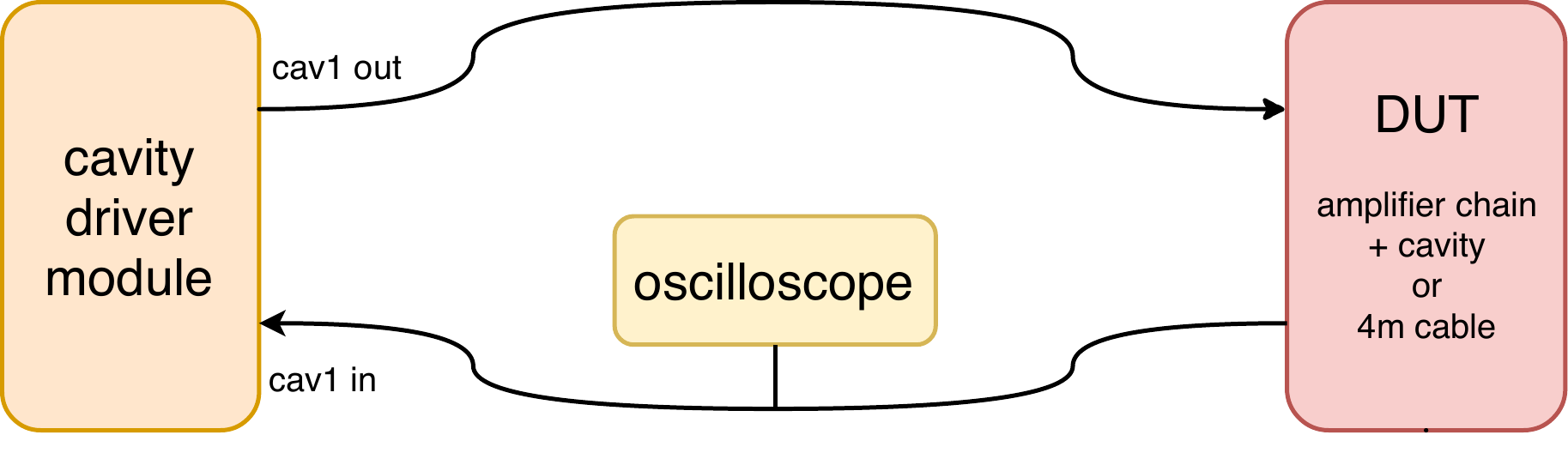} 
 \caption{Test setup of the cavity driver module.}
\label{fig:meas_setup}
\end{figure}

\begin{figure}[t] 
\centering
\includegraphics[width=\linewidth]{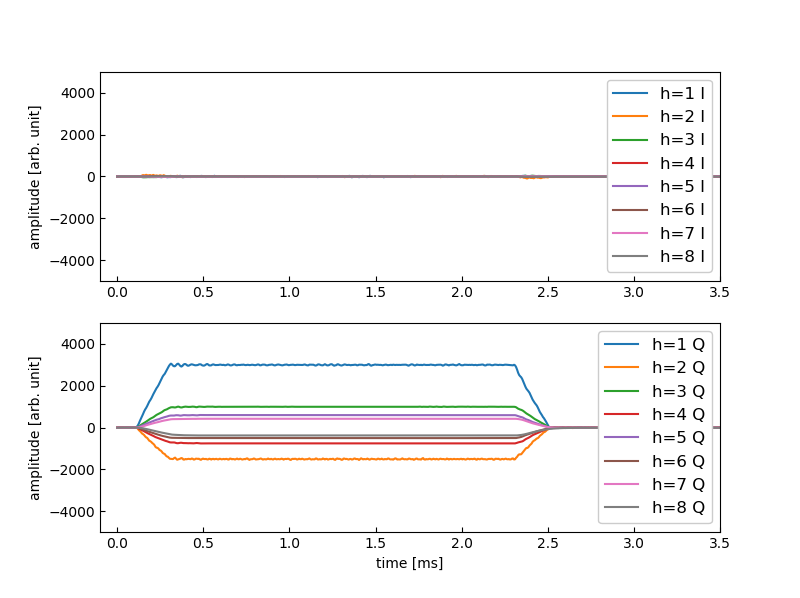} 
 \caption{Measured I/Q amplitudes of eight harmonics ($h=1..8$).}
\label{fig:iq_measured}
\end{figure}

\begin{figure}[t] 
\centering
\includegraphics[width=\linewidth]{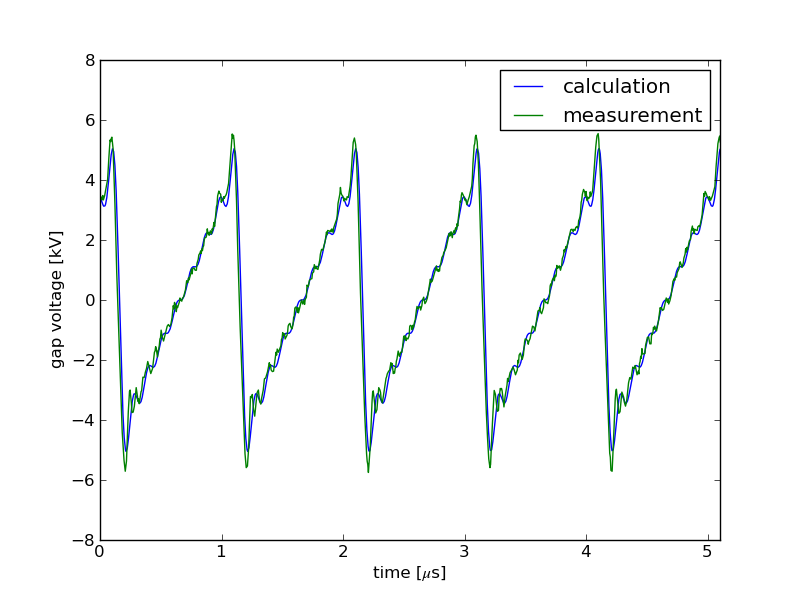} 
 \caption{Comparison of the measured and calculated cavity gap voltage waveforms.}
\label{fig:compare_waveforms}
\end{figure}

\section{Preliminary test results}
\subsection{Multiharmonic vector rf control}
The test setup of the cavity driver module is shown in Fig.~\ref{fig:meas_setup}.
The rf output for the cavity 1 is led to the DUT (device under test) and
the output of the DUT is fed into the cavity 1 input of the driver module.
The DUT is the amplifier chain and the cavity.
The phase offset LUT was set so that the feedback loop can be closed.

To demonstrate the performance of the multiharmonic vector rf control,
a sawtooth wave is generated. The Fourier series $f(t)$ of a sawtooth
wave with a frequency $f_1$ and the amplitude 1 up to $m$-th harmonic is
\begin{align}
 f(t) &= \frac{2}{\pi} \sum_{h=1}^m  \frac{(-1)^{h+1}}{h} \sin 2\pi h f_1 t, \label{eq:fourier}
\end{align}
where $h$ is the harmonic number. The module can control eight harmonics ($h=1..8$). 
In the test, $f_1$ was set to 1~MHz and the I/Q amplitude of the revolutional
harmonic ($h=1$) is set to (0,3000), which are digital values.
The amplitudes of the higher harmonics are set according to (\ref{eq:fourier}).

The measured I/Q signals of the eight harmonics ($h=1..8$) 
and the comparison of the measured and calculated cavity gap voltage waveforms
are plotted in Fig.~\ref{fig:iq_measured} and Fig.~\ref{fig:compare_waveforms},
respectively.
One can see that the I/Q amplitudes of the harmonics are very close to the set points.
The measured and calculated waveforms nicely agree.

The performance of the multiharmonic vector rf control is promising.
The beam loading compensation up to $h=8$ with the vector rf control
is foreseen. 
Also, the third ($h=6$) and fourth ($h=8$) harmonic voltages in addition
to the existing dual harmonic operation may improve the performance
of the bunch shaping to alleviate the space charge
effects.

\subsection{High speed serial communication and vector sum function}
To examine the high speed serial communication and vector sum function,
an I/Q signal rotated by a phase $\theta$
for the selected harmonic ($h=1$) is sent from the cavity
driver module to the high speed serial communication module.
The normalized vector sum I/Q signal is sent to the common function
module. A 4~m long cable is used as the DUT for this test.

\begin{figure}[t] 
\centering
\includegraphics[width=\linewidth]{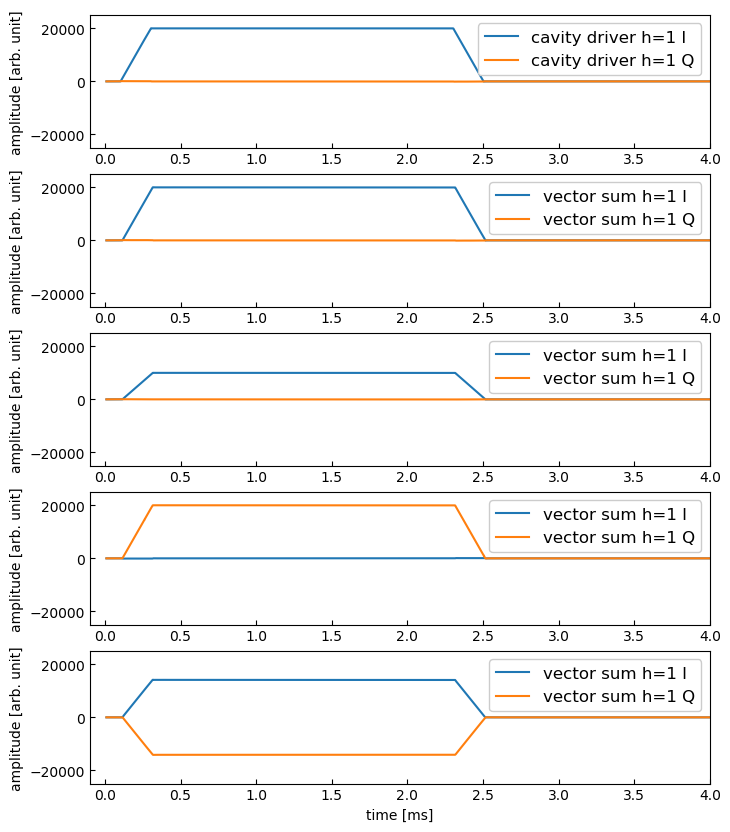} 
 \caption{I/Q waveforms. The top is the I/Q signal measured by the cavity driver.
The others are received vector sum signals at the common function module. 
From the second top to the bottom,
the signal normalized by 1 without rotation, the signal normalized by factor of 2 without rotation,
the signal normalized by 1 with the rotation angle of 90 degrees, 
and the signal normalized by 1 with the rotation angle of $-45$ degrees.}
\label{fig:IQ_vector_sum_wfm}
\end{figure}

The top plot of Fig.~\ref{fig:IQ_vector_sum_wfm} shows the measured
I/Q signal at the cavity driver module. The amplitude of I is 20000
and the Q is zero. The envelope is a trapezoid. The rising and falling time
is 0.2~ms and the flattop width is 2~ms.

The other plots in Fig.~\ref{fig:IQ_vector_sum_wfm} are 
received vector sum signals at the common function module.
The second top plot shows the signal normalized by 1 without
rotation. It is identical to the I/Q signal measured by the cavity
driver module. The middle plot shows
the signal normalized by factor of 2 without rotation. The amplitude
is a half of the original signal. 

The second bottom plot show the signal normalized by 1 with the rotation angle of 90 degrees.
The amplitude of I is zero and the Q is 20000. The bottom is the 
signal normalized by 1 with the rotation angle of $-45$ degrees.
The amplitude of I is close to $20000 \times 1/\sqrt{2}=14142$ and the Q is 
similar to I with negative sign.

This simple test proves that the vector sum function works correctly as designed.
We should note that we did not see any errors on the I/Q waveforms;
the high speed serial transfer via the Xilinx Aurora is very stable.

\section{Summary and outlook}
We summarize the article as follows.

The existing LLRF control system has been working nicely without major
problems for more than ten yeas.  However, it will be difficult to
maintain the system in near future because of the discontinued FPGAs on
the system.  The MicroTCA.4 based next generation LLRF control system is
now under development. 

Similar LLRF functions are to be implemented in
the new system with several new features.
The key feature of the new system is the multiharmonic vector rf control,
which would compensate the heavy beam loading in the wideband rf cavity
and can expand the performance of the longitudinal painting injection.
With the capability of the MicroTCA.4 backplane for high speed serial communication,
the sophisticated signal transfer between the modules is realized.

We will add five cavity driver modules to control twelve cavities
and implement the remaining LLRF functions.
We plan to replace the existing system with the new system during
the summer maintenance period in 2019. Prior to the replacement,
we will perform beam tests with the new system mainly focused on
the beam loading compensation.

\section*{Acknowledgments}
We would like to thank Heiko Damerau and John Molendijk for fruitful
discussions on LLRF topics. We also would like to thank the J-PARC
writing support group, which continuously encouraged us to write up this
article. Finally, we would like to thank all the members of the J-PARC.

\bibliographystyle{IEEEtran}
\bibliography{tamura-bib_eng_only}

\end{document}